\documentclass[amsmath,amssymb,superscriptaddress,nobalancelastpage,aps,prl,twocolumn]{revtex4-2}

\usepackage{graphicx}
\usepackage{varioref}
\usepackage{xr-hyper}
\usepackage{xcolor}
\usepackage{nicefrac}
\usepackage{xfrac}
\usepackage{hyperref}
\hypersetup{colorlinks,linkcolor=blue,urlcolor=blue,citecolor=blue}
\usepackage{ulem}
\usepackage{lineno}
\usepackage{amsmath} 
\usepackage{amssymb}
\usepackage{physics}
\usepackage{soul}

\begin{document}

\title{Unconventional transverse transport above and below the magnetic transition temperature in Weyl semimetal EuCd$_2$As$_2$}

\author{Y. Xu}
\email{yangxu@physik.uzh.ch}
\affiliation{Physik-Institut, Universit\"{a}t Z\"{u}rich, Winterthurerstrasse 190, CH-8057 Z\"{u}rich, Switzerland}

\author{L. Das}
\affiliation{Physik-Institut, Universit\"{a}t Z\"{u}rich, Winterthurerstrasse 190, CH-8057 Z\"{u}rich, Switzerland}

\author{J. Z. Ma}
\affiliation{Department of Physics, City University of Hong Kong, Kowloon, Hong Kong}
\affiliation{Swiss Light Source, Paul Scherrer Institut, Villigen CH-5232, Switzerland}

\author{C. J. Yi}
\affiliation{Beijing National Laboratory for Condensed Matter Physics, Institute of Physics, Chinese Academy of Sciences, Beijing 100190, China}
\affiliation{University of Chinese Academy of Sciences, Beijing 100049, China}

\author{S. M. Nie}
\affiliation{Department of Materials Science and Engineering, Stanford University, Stanford, CA 94035, USA}

\author{Y. G. Shi}
\affiliation{Beijing National Laboratory for Condensed Matter Physics, Institute of Physics, Chinese Academy of Sciences, Beijing 100190, China}

\author{A. Tiwari}
\affiliation{Physik-Institut, Universit\"{a}t Z\"{u}rich, Winterthurerstrasse 190, CH-8057 Z\"{u}rich, Switzerland}
\affiliation{Condensed Matter Theory Group, Paul Scherrer Institute, CH-5232 Villigen PSI, Switzerland}

\author{S. S. Tsirkin}
\affiliation{Physik-Institut, Universit\"{a}t Z\"{u}rich, Winterthurerstrasse 190, CH-8057 Z\"{u}rich, Switzerland}

\author{T. Neupert}
\affiliation{Physik-Institut, Universit\"{a}t Z\"{u}rich, Winterthurerstrasse 190, CH-8057 Z\"{u}rich, Switzerland}

\author{M. Medarde}
\affiliation{Laboratory for Multiscale Materials Experiments, Paul Scherrer Institut, CH-5232 Villigen PSI, Switzerland}

\author{M. Shi}
\affiliation{Swiss Light Source, Paul Scherrer Institut, Villigen CH-5232, Switzerland}

\author{J. Chang}
\email{johan.chang@physik.uzh.ch}
\affiliation{Physik-Institut, Universit\"{a}t Z\"{u}rich, Winterthurerstrasse 190, CH-8057 Z\"{u}rich, Switzerland}

\author{T. Shang}
\email{tshang@phy.ecnu.edu.cn}
\affiliation{Key Laboratory of Polar Materials and Devices (MOE), School of Physics and Electronic Science, East China Normal University, Shanghai 200241, China}
\affiliation{Physik-Institut, Universit\"{a}t Z\"{u}rich, Winterthurerstrasse 190, CH-8057 Z\"{u}rich, Switzerland}
\affiliation{Laboratory for Multiscale Materials Experiments, Paul Scherrer Institut, CH-5232 Villigen PSI, Switzerland}

\date{\today}

\begin{abstract}
As exemplified by the growing interest in the quantum anomalous Hall effect, the research on topology as an organizing principle of quantum matter is greatly enriched from the interplay with magnetism. In this vein, we present a combined electrical and thermoelectrical transport study on the magnetic Weyl semimetal EuCd$_2$As$_2$. Unconventional contribution to the anomalous Hall and anomalous Nernst effects were observed both above and below the magnetic transition temperature of EuCd$_2$As$_2$, indicating the existence of significant Berry curvature. EuCd$_2$As$_2$ represents a rare case in which this unconventional transverse transport emerges both above and below the magnetic transition temperature in the same material. The transport properties evolve with temperature and field in the antiferromagnetic phase in a different manner than in the paramagnetic phase, suggesting different mechanisms to their origin. Our results indicate EuCd$_2$As$_2$ is a fertile playground for investigating the interplay between magnetism and topology, and potentially a plethora of topologically nontrivial phases rooted in this interplay.

\end{abstract}


\maketitle
The last few decades have witnessed the effort of researchers seeking unique insights into the physics of materials from the view of topology. When magnetism, the most time-honored branch in condensed matter physics, is incorporated into this view, novel phenomena are to be expected. One notable manifestation is an additional contribution to transverse transport properties. For example, the topological Hall effect has been intensively studied in systems exhibiting non-coplanar spin textures with finite scalar spin chirality $\chi_{ijk}~=~\mathbf{S_i}\cdot (\mathbf{S_j}\times \mathbf{S_k})$ (where $\mathbf{S_n}$ are the spins), such as skyrmions~\cite{neubauer_topological_2009,gayles_dzyaloshinskii-moriya_2015,lee_unusual_2009,kanazawa_large_2011,li_robust_2013,franz_real-space_2014,huang_extended_2012,schulz_emergent_2012,qin_emergence_2019,matsuno_interface-driven_2016,kurumaji_skyrmion_2019}, hedgehogs~\cite{kanazawa_critical_2016,fujishiro_topological_2019}, hopfions~\cite{gobel_topological_2020}, merons~\cite{puphal_topological_2020}, and magnetic bubbles~\cite{vistoli_giant_2019}. The topological Hall effect is an unconventional contribution to the anomalous Hall effect (UAHE) in addition to the conventional component that scales with magnetization~\cite{neubauer_topological_2009,gayles_dzyaloshinskii-moriya_2015,lee_unusual_2009,kanazawa_large_2011,li_robust_2013,franz_real-space_2014,huang_extended_2012,schulz_emergent_2012,qin_emergence_2019,matsuno_interface-driven_2016,kurumaji_skyrmion_2019}. Alternative to this real-space scenario, UAHE has also been observed in systems with band structure anomalies such as Weyl points near the Fermi level, which carry significant Berry curvature that acts like an effective magnetic field in the momentum space~\cite{onoda_anomalous_2004,nakatsuji_large_2015,liang_ultrahigh_2015,liang_anomalous_2017,liang_anomalous_2018,suzuki_large_2016}. 

It has been demonstrated recently that the antiferromagnet EuCd$_2$As$_2$ with $T_{\rm N} \sim$ 9.5 K would be an ideal candidate for the study of the interplay between magnetism and topology, as it exhibits various topological states in both the antiferromagnetic (AFM) and paramagnetic (PM) phases~\cite{hua_dirac_2018,ma_spin_2019,soh_ideal_2019,wang_single_2019,ma_emergence_2020,jo_manipulating_2020}. In the AFM phase, depending on the direction of the Eu magnetic moments, various nontrivial topological ground states have been predicted, such as magnetic topological Dirac semimetal, or axion insulator, AFM topological crystalline insulator, and higher order topological insulator~\cite{ma_emergence_2020}. When the spins are aligned along the $c$ axis by external magnetic field, a single pair of Weyl points appears near the Fermi level~\cite{ma_emergence_2020,soh_ideal_2019}. In the PM phase, EuCd$_2$As$_2$ turns out to be the first discovered centrosymmetric Weyl semimetal where ferromagnetic (FM) spin fluctuations, instead of long-range magnetic order, lift the Kramers degeneracy~\cite{ma_spin_2019}. A rich magnetic phase diagram is thus expected for EuCd$_2$As$_2$~\cite{ma_emergence_2020,soh_ideal_2019,wang_single_2019,hua_dirac_2018}. Profound insights into the interplay between magnetism and topology, from the exploration of this phase diagram in the context of unconventional transport, can be reasonably foreseen and, thus, such a transport study is highly desired.

Thermoelectrical transport can provide additional information than electrical transport, as it is usually more sensitive to the Berry curvature near the Fermi level~\cite{behnia_nernst_2016}. As a thermoelectrical counterpart of the UAHE, the unconventional contribution to the anomalous Nernst effect (UANE) has only been reported in few systems~\cite{ikhlas_large_2017,hirschberger_topological_2019,liang_anomalous_2017,liang_anomalous_2018,shiomi_topological_2013}. Linked to each other by the Mott relation~\cite{hirschberger_topological_2019}, the observation of a large UAHE, however, does not guarantee a large UANE~\cite{xiao_berry_2010,ikhlas_large_2017}. Here, we present a systematic study of the electrical and thermoelectrical transport properties of EuCd$_2$As$_2$. The presence of finite UAHE and UANE above $T_{\rm N}$ in EuCd$_2$As$_2$ represents a rare case of unconventional transverse transport beyond the ordinary and conventional anomalous contribution. Above $T_{\rm N}$, both positive and negative UAHE and UANE were observed and attributed to the fluctuating Weyl points near the Fermi level. Below $T_{\rm N}$, the UAHE and UANE evolve with field and temperature in a qualitatively different manner compared to that above $T_{\rm N}$, and their origins may be attributed to a concerted effort from the real-space and momentum-space scenarios contributing to the Berry curvature.

The zero-field N\'eel temperature of our EuCd$_2$As$_2$ single crystal is $T_{\rm N} \sim$ 9.5 K (see Fig. S1 in the Supplemental Material~\cite{SM_note}), consistent with previous reports~\cite{ma_spin_2019,soh_ideal_2019,ma_emergence_2020,jo_manipulating_2020}. The magnetic field dependence of the magnetization $M$, longitudinal resistivity $\rho_{xx}$, Hall resistivity $\rho_{xy}$, and Nernst signal $S_{xy}$ of EuCd$_2$As$_2$ for different temperatures are shown in Fig 1. Anomalies in the low field region are already evident in these isotherms without further analysis. For transverse transport $\rho_{xy}$ and $S_{xy}$, the most prominent feature is the presence of low-field peaks superimposed on a (quasi-) linear background. Both the associated field scale and amplitude of the peaks exhibit a strong variation with temperature. For each isotherm, the empirical relation $\rho_{xy}$ = $\rho_{xy}^O$ + $\rho_{xy}^A$ = $\rho_{xy}^O$ + $\rho_{xy}^{\rm CA}$ + $\rho_{xy}^{\rm UA}$ = $R_0 \mu_0 H$ + $S_H \rho_{xx}^2 M$ + $\rho_{xy}^{\rm UA}$ were applied to separate the different contributions to $\rho_{xy}$. This decomposition procedure is shown in Fig. S2~\cite{SM_note}. The additivity of the above relation holds for $\rho_{xy} \ll$ $\rho_{xx}$~\cite{kanazawa_large_2011}, which is the case here in EuCd$_2$As$_2$ [see Figs. 1(b) and (c)]. Here, $R_0$ and $S_H$ are constants for a given isotherm. $\rho_{xy}^O$ and $\rho_{xy}^A$ are the ordinary and anomalous Hall contribution, respectively. $\rho_{xy}^A$ can be further divided into the conventional anomalous term $\rho_{xy}^{\rm CA}$ and an unconventional term $\rho_{xy}^{\rm UA}$, respectively. The form of $\rho_{xy}^{\rm CA}$ used here assumes the domination of the intrinsic mechanism for the conventional anomalous Hall effect~\cite{lee_hidden_2007,nagaosa_anomalous_2010}. Assuming instead a skew scattering dominated $\rho_{xy}^{\rm CA}$ barely affects the results (see Sec. II in the Supplemental Material~\cite{SM_note}). Isotherms of the UAHE term $\rho_{xy}^{\rm UA}$ derived from the above decomposition procedure are shown in Fig. 2(a). 

The Nernst conductivity $\alpha_{xy}$ is obtained using $\alpha_{xy}$ = $\sigma_{xx}S_{xy}$ + $\sigma_{xy}S_{xx}$ + $\sigma_{xx} \kappa_H / \kappa$~\cite{shiomi_topological_2013,onose_anomalous_2007}, where $\sigma_{xx}$ and $\sigma_{xy}$ are the longitudinal and Hall conductivity, respectively, $S_{xx}$ is the Seeback signal (see Fig. S1~\cite{SM_note}), $\kappa_H $ is the thermal Hall, and $\kappa$ is the thermal conductivity (see Sec. III in the Supplemental Material~\cite{SM_note}). We employ similar procedure as that for $\rho_{xy}$ to $\alpha_{xy}$~\cite{shiomi_topological_2013}: $\alpha_{xy}$ = $\alpha_{xy}^O$ + $\alpha_{xy}^A$ = $\alpha_{xy}^O$ + $\alpha_{xy}^{\rm CA}$ + $\alpha_{xy}^{\rm UA}$ = $Q_0 \mu_0 H$ + $Q_s M$ + $\alpha_{xy}^{\rm UA}$, with $Q_0$ and $Q_s$ being constants for a specific isotherm. The resulting isotherms of the UANE term $\alpha_{xy}^{\rm UA}$ are shown in Fig. 2(b). The temperature dependence of the amplitude of the positive and negative peaks of $\rho_{xy}^{\rm UA}$ and $\alpha_{xy}^{\rm UA}$ are summarized in Fig. 3(a). The peak positions at different temperatures are superimposed on the contour plot of $\rho_{xy}^{\rm UA}$ in Fig. 3(b), forming a rich phase diagram covering both the AFM and PM phase of EuCd$_2$As$_2$. Apart from a bifurcation of peak positions at low temperatures [Fig. 3(b)], the electrical and thermoelectrical transport properties are congruent and, thus, corroborate each other.

\begin{figure}
\begin{center}
 		\includegraphics[width=0.42\textwidth]{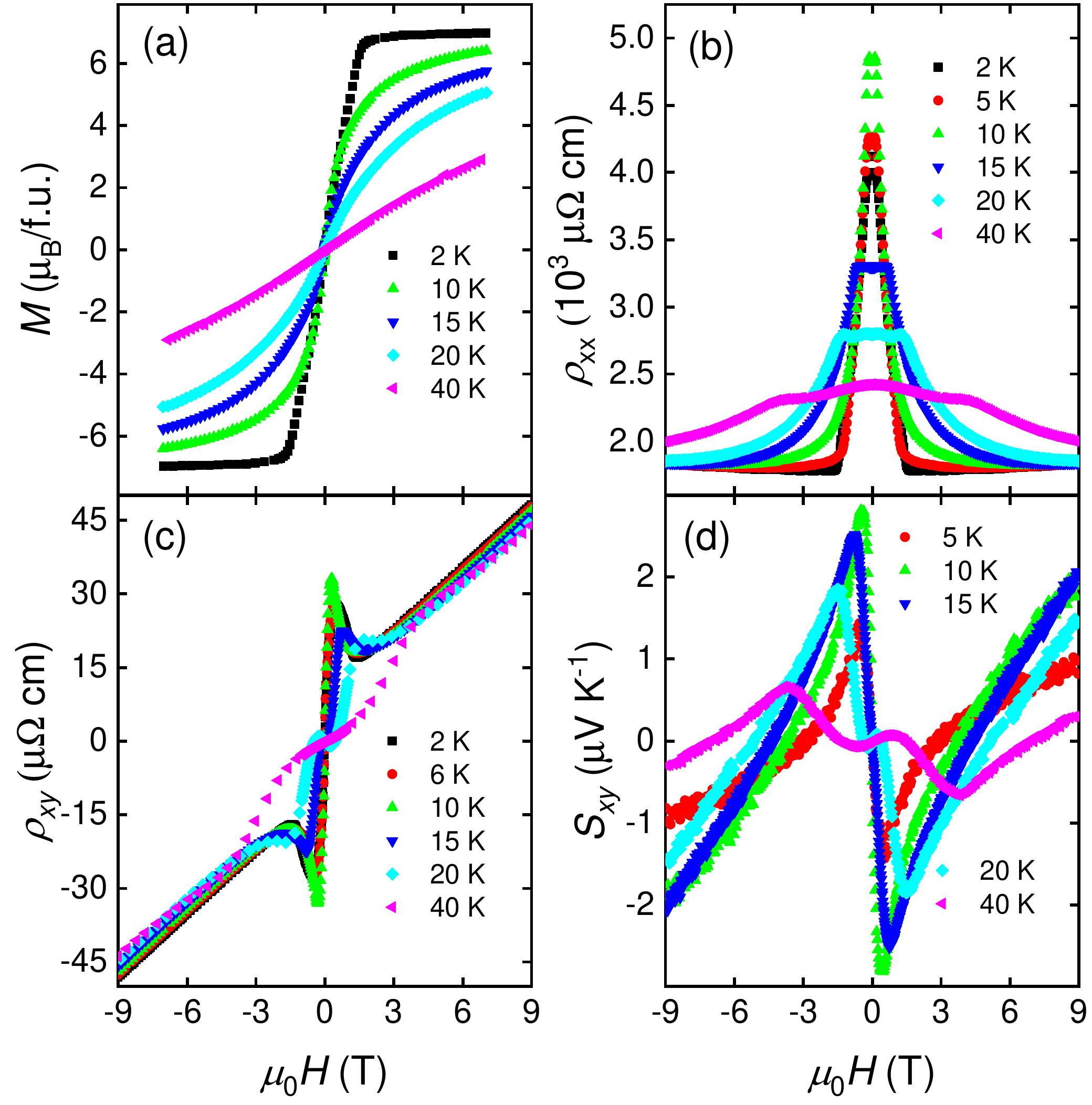}
 	\end{center}
\centering
\caption{The magnetic field dependence of the (a) magnetization $M$, (b) resistivity $\rho_{xx}$, (c) Hall resistivity $\rho_{xy}$, and (d) Nernst signal $S_{xy}$ of EuCd$_2$As$_2$ for selected temperatures. The $\rho_{xx}$ isotherms are symmetrized while the $\rho_{xy}$ and $S_{xy}$ isotherms antisymmetrized. Magnetic fields $\mu_0 H$ ($\mu_0$ being the vacuum permeability) up to 9 T were applied along the $c$ axis, while electrical and thermal currents, and the measured voltage drops were all in the $ab$ plane. More information on the measurement protocol can be found in Sec. I in the Supplemental Material~\cite{SM_note}.}
\end{figure}

\begin{figure}
\begin{center}
 		\includegraphics[width=0.39\textwidth]{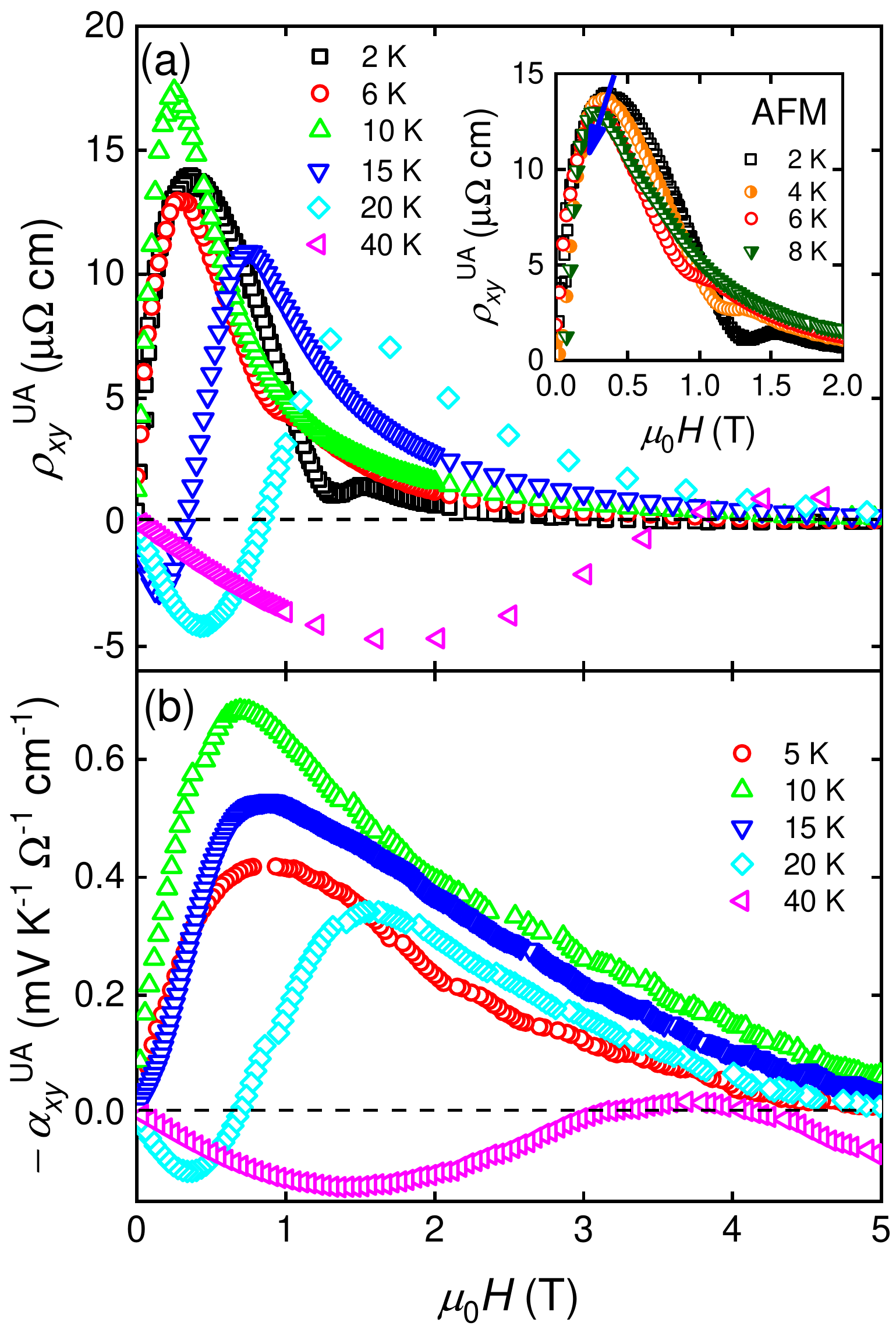}
 	\end{center}
\caption{(a,b) The magnetic field dependence of the unconventional contribution to the anomalous Hall resistivity $\rho_{xy}^{\rm UA}$ and the the unconventional contribution to the anomalous Nernst conductivity $-\alpha_{xy}^{\rm UA}$ for selected temperatures. The isotherms are shown for $\mu_0H \leq$ 5 T, since they are basically featureless at higher fields. The inset of (a) shows an enlarged view of the positive peak on the $\rho_{xy}^{\rm UA}$ isotherms in the antiferromagnetic phase. The blue arrow tracks the evolution of the peak with increasing temperature.}
\end{figure}

Considering the observations in Figs. 2 and 3(a)-(b), it is obvious that both $\rho_{xy}^{\rm UA}$ and $\alpha_{xy}^{\rm UA}$ exhibit a two-stage behavior, $i.e.$, they evolve in a qualitatively different manner in the AFM phase compared to in the PM phase: (i) The field scale associated with the positive peak decreases with increasing temperature in the AFM phase while it increases in the PM phase. As a result, the field scale of the 10 K (slightly above $T_{\rm N}$) isotherm is the smallest among all the isotherms. (ii) The width of the positive peak decreases with increasing temperature in the AFM phase while it increases in the PM phase. (iii) The amplitude of the positive peak decreases with increasing temperature both in the AFM and PM phase, but there is a sudden increase around $T_{\rm N}$, such that the 10 K isotherm exhibits a peak amplitude larger than any of the isotherms in the AFM phase. (iv) The negative peak is only observed in the PM phase. As temperature increases in the PM phase, the positive peak gradually gives way to a negative peak. All these observations suggest different origins of the UAHE and UANE below and above $T_{\rm N}$.

To our knowledge, there is no previous report of anomalous Nernst effect above the magnetic transition temperature in any material. On the other hand, the anomalous Hall effect above the magnetic transition temperature has only been observed recently in a few systems, $e.g.$, GdPtBi and YbPtBi~\cite{suzuki_large_2016,guo_evidence_2018}. However, there the anomalous Hall signal above $T_{\rm N}$ is likely a vestige of the signal below $T_{\rm N}$, with no anomaly around $T_{\rm N}$, pointing to a similar underlying mechanism at all temperatures~\cite{suzuki_large_2016,guo_evidence_2018}, as opposed to the two-stage behavior in EuCd$_2$As$_2$. These facts endow EuCd$_2$As$_2$ a unique place among magnetic topological materials and form the main finding of this work.

\textit{UAHE and UANE above $T_N$}: Finite UAHE and UAHE above $T_{\rm N}$ can only be attributed to momentum-space Berry curvature anomalies, such as Weyl points~\cite{realinPM_note}. As a source of Berry curvature, Weyl points have been argued to underlie the anomalous Hall and anomalous Nernst effect in a wide range of ferromagnets~\cite{liu_giant_2018,yang_giant_2020,sakai_giant_2018,guin_anomalous_2019}, antiferromagnets~\cite{nakatsuji_large_2015,ikhlas_large_2017,suzuki_large_2016,guo_evidence_2018,nayak_large_2016}, and nonmagnetic topological semimetals~\cite{liang_ultrahigh_2015,liang_anomalous_2017,liang_anomalous_2018,caglieris_anomalous_2018,watzman_dirac_2018}. Indeed, Weyl points in the PM phase of EuCd$_2$As$_2$ were observed by angle-resolved photoemission spectroscopy (ARPES)~\cite{ma_spin_2019}. The theoretical modelling for the ARPES spectra is such that at any given time, the system can be divided into FM correlation domains~\cite{ma_spin_2019}. The direction of the magnetic moments is the same within each domain, but is random and uniformly distributed in different domains~\cite{ma_spin_2019}. This is illustrated in Fig. 3(c): FM domains with moments along all directions give rise to a broadened band structure, and a distribution of Weyl points on and off the $\Gamma - A$ high symmetry line. In each FM domain, Weyl points emerge due to the breaking of time-reversal symmetry. Macroscopically, however, the polarity of each Weyl points pair is compensated by the pair in another domain with opposite magnetization. Consequently, the positive and negative contribution to the transverse transport is cancelled out. This well explains the lack of a spontaneous component of the UAHE and UANE above $T_{\rm N}$. When applying external magnetic field along the $c$ axis, domains with moments along $c$ is more favored, which translates to a higher probability for the Weyl points on the $\Gamma - A$ line [Fig. 3(d)]. The non-zero net polarity of all the Weyl points pairs gives rise to finite UAHE and UANE.

Another notable feature of the UAHE and UANE above $T_{\rm N}$ in EuCd$_2$As$_2$ is the coexistence of positive and negative values. This is attributed to the influence by Weyl points pairs farther from the Fermi level, due to the thermal broadening of the Fermi surface (see Sec. VI in the Supplemental Material~\cite{SM_note}), and is consistent with the Weyl-points-induced UAHE and UANE scenario. Previous reports on anomalous transverse transport of opposite signs are limited to the topological Hall effect induced by a real-space scenario below the magnetic transition temperature~\cite{kanazawa_large_2011,kanazawa_critical_2016,fujishiro_topological_2019,ishizuka_spin_2018}.

\begin{figure*}
\begin{center}
 		\includegraphics[width=0.95\textwidth]{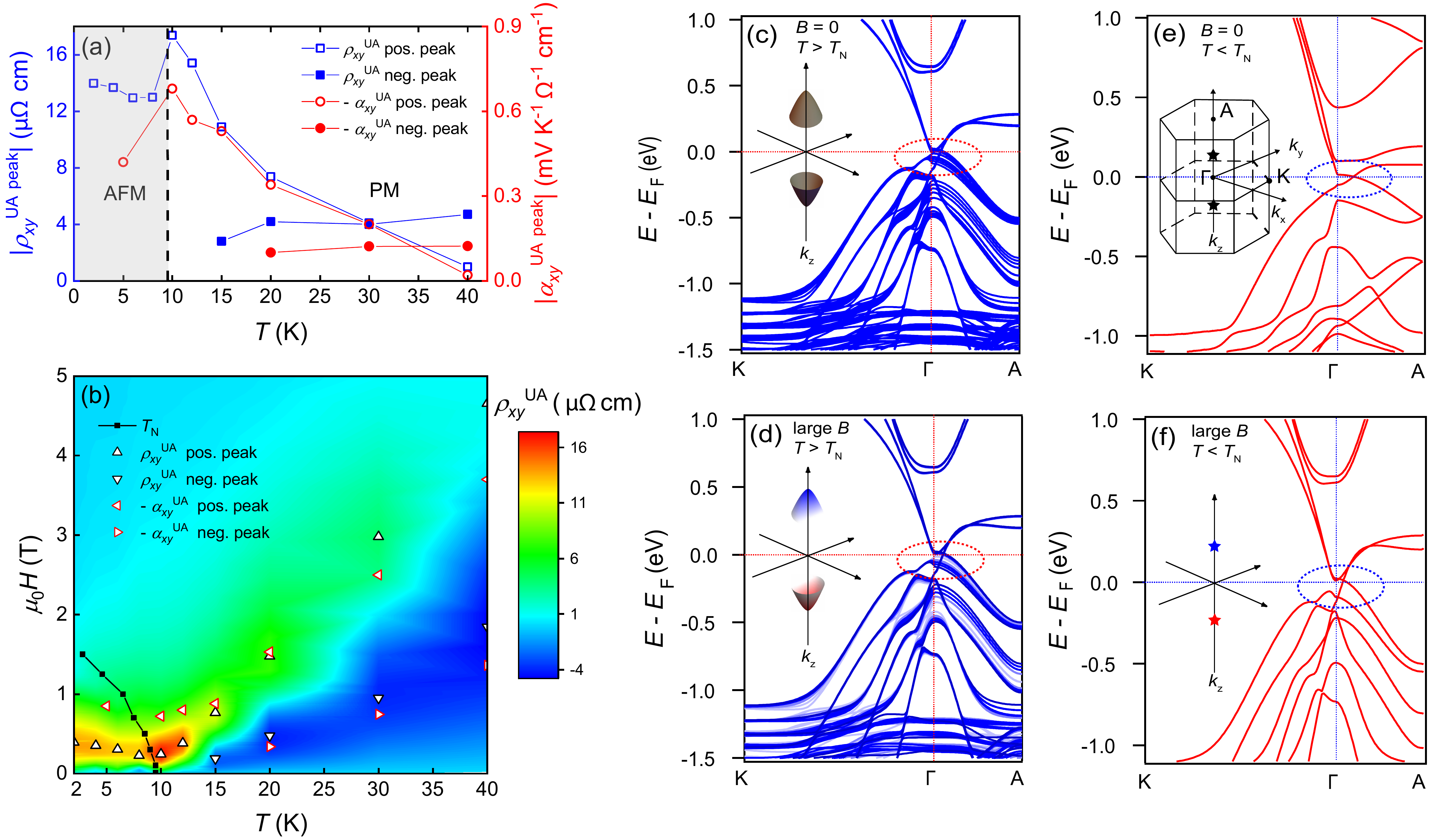}
 	\end{center}
\centering

\caption{
(a) Temperature dependence of the absolute values of $\rho_{xy}^{\rm UA}$ (blue squares, left axis) and $\alpha_{xy}^{\rm UA}$ (red circles, right axis) at the positive peak (open symbols) and negative peak (filled symbols). The vertical dashed line separates the antiferromagnetic and the paramagnetic phase. (b) The $H$-$T$ (magnetic field versus temperature) phase diagram of EuCd$_2$As$_2$. The colors represent the magnitude of the unconventional contribution to the anomalous Hall resistivity $\rho_{xy}^{\rm UA}$. The N\'eel temperatures $T_{\rm N}$ (filled squares) are extracted from susceptibility measurements (see Fig. S1~\cite{SM_note}). The positions of the positive and negative peaks of $\rho_{xy}^{\rm UA}$ (upper and lower triangles) and $-\alpha_{xy}^{\rm UA}$ (left and right triangles) are also superimposed. (c)-(f) Electronic band structure of EuCd$_2$As$_2$ calculated for temperatures above and below $T_{\rm N}$ and at zero and large external magnetic field, adapted from Refs.~\cite{ma_spin_2019,ma_emergence_2020}. The inset in (c)-(f) illustrates the position of the Weyl (Dirac) points closest to the Fermi level in the Brillouin zone. The cones in (c) and (d) depict all possible positions of the Weyl points, which originate from fluctuating FM domains. The pentagrams in (e) and (f) depict the position of the Dirac points and Weyl points, respectively. The gray and black color for the cones and pentagrams indicate a zero net topological charge, while the blue and red color indicate a $-$1 and +1 topological charge, respectively. In (d), the opacity gradient for the band dispersion indicates the spectral weight, while for the cones it indicates the probability of Weyl points at a specific position in the Brillouin zone.}

\end{figure*}

\textit{UAHE and UANE below $T_N$}: It was proposed that EuCd$_2$As$_2$ exhibits an A-type AFM structure, $i.e.$, the spins form ferromagnetic layers which stack antiferromagnetically along the $c$ axis~\cite{rahn_coupling_2018,wang_anisotropic_2016,schellenberg_121sb_2011}. At zero field, the moments lie in the $ab$ plane, and there is a pair of Dirac points along $\Gamma - A$ around the Fermi level~\cite{Dirac_note,ma_emergence_2020} [see Fig. 3(e)]. The Dirac points split into several pairs of Weyl points as external magnetic field increases, which start to contribute to the UAHE and UANE. Finally, at large field, a single pair of Weyl points along $\Gamma - A$ is left, as shown in Fig. 3(f). More discussions on the field evolution of the UAHE and UANE are provided in Sec. VI in the Supplemental Material~\cite{SM_note}.

We have demonstrated that the Berry curvature associated with the Weyl points contributes to the UAHE and UANE in EuCd$_2$As$_2$, both for temperatures above and below $T_{\rm N}$, with two different physical pictures [Figs. 3(c)-(f)]. This is consistent with the two-stage behavior of the UAHE and UANE. On the other hand, a real-space scenario, which is relevant only below $T_{\rm N}$, acts as another possible source for the two-stage behavior. In fact, the AFM portion of the phase diagram shown in Fig. 3(b) strongly resembles that of systems with a real-space scenario induced UAHE, $i.e.$, a topological Hall effect. For instance, in Gd$_2$PdSi$_3$, a skyrmion lattice phase is sandwiched by two incommensurate spin-state phases, and a finite $\rho_{xy}^{\rm UA}$ can only be observed in the skyrmion lattice phase~\cite{kurumaji_skyrmion_2019}. The $\rho_{xy}^{\rm UA}$ peak also exhibits an amplitude, field scale, and width that decrease with increasing temperature~\cite{kurumaji_skyrmion_2019}, as is the case in the AFM phase of EuCd$_2$As$_2$. Based on the current data, a contribution from a real-space scenario can neither be pinned down nor excluded in the AFM phase of EuCd$_2$As$_2$~\cite{magnetization_note}. The genuine magnetic structure of EuCd$_2$As$_2$ under magnetic field could be more complex, considering the sizable frustration in this material~\cite{ma_spin_2019,neutron_note}. Real-space probes such as Lorentz transmission electron microscopy are required to search for potential spin textures with finite chirality. These nontrivial spin textures have been observed mostly in systems with a noncentrosymmetric crystal structure~\cite{neubauer_topological_2009,gayles_dzyaloshinskii-moriya_2015,lee_unusual_2009,kanazawa_large_2011,li_robust_2013,franz_real-space_2014,huang_extended_2012,schulz_emergent_2012,qin_emergence_2019,matsuno_interface-driven_2016}, as compared to the centrosymmetric EuCd$_2$As$_2$. As evidenced by the recent upsurge of research interest in centrosymmetric skyrmion systems~\cite{kurumaji_skyrmion_2019,hirschberger_skyrmion_2019,li_large_2019,khanh_nanometric_2020}, the existence of nontrivial spin textures in the AFM phase of EuCd$_2$As$_2$, would be an interesting possibility to test.

Finally we compare the magnitude of the UAHE and UANE in EuCd$_2$As$_2$ with other systems. In a real-space scenario, the $\lvert\rho_{xy}^{\rm UA}\rvert${$\rm^{max}$} $\sim$ 20 $\mu \Omega$~cm reported here in EuCd$_2$As$_2$ is significant, as compared to, $e.g.$, the so-called giant topological Hall effect in Gd$_2$PdSi$_3$ with $\lvert\rho_{xy}^{\rm UA}\rvert${$\rm^{max}$} $\sim$ 3 $\mu \Omega$~cm~\cite{kurumaji_skyrmion_2019}. This is consistent with the fact that Weyl points contribute to $\rho_{xy}^{\rm UA}$ at temperatures both above and below $T_{\rm N}$. In a momentum-space scenario, it is more appropriate to compare the magnitude of $\sigma_{xy}^{\rm UA}$ and the Hall angle $\Theta_H^{\rm UA}$ = $\sigma_{xy}^{\rm UA}/ \sigma_{xx}^{\rm UA}$~\cite{onoda_anomalous_2004}. The $\lvert\sigma_{xy}^{\rm UA}\rvert${$\rm^{max}$} $\sim$ 1 $\Omega^{-1}$~cm$^{-1}$ and $\lvert\Theta_H^{\rm UA}\rvert${$\rm^{max}$} $\sim$ 0.01 (see Sec. IV in the Supplemental Material~\cite{SM_note}) reported here in EuCd$_2$As$_2$ are comparable to the values estimated from a previous study~\cite{soh_ideal_2019}, and are one order of magnitude smaller than that in GdPtBi~\cite{suzuki_large_2016}. It is worth noticing that the UANE in EuCd$_2$As$_2$ is larger than the Weyl-points-induced UANE in Mn$_3$Sn~\cite{ikhlas_large_2017} (see Table S1 in the Supplemental Material~\cite{SM_note}).  

In summary, our electrical and thermoelectrical transport measurements on EuCd$_2$As$_2$ produce highly consistent results: unconventional contribution to the anomalous Hall effect and anomalous Nernst effect is revealed both below and above the antiferromagnetic ordering temperature $T_{\rm N}$. The unconventional term is attributed to the Weyl points near the Fermi level, both below and above $T_{\rm N}$, although in the former case a contribution from topological real-space spin textures cannot be excluded. The existence of an unconventional term above $T_{\rm N}$ is in itself uncommon. Moreover, the two-stage evolution of the unconventional term in the antiferromagnetic and in the paramagnetic phase hints at their different mechanism originating from the interplay of magnetism and topology. The traversal among the topologically nontrivial phases of EuCd$_2$As$_2$ may also be achieved by symmetry-breaking perturbations other than magnetic field~\cite{watzman_dirac_2018,niu_quantum_2019,sanjeewa_evidence_2020,QSH_note}. 

\begin{acknowledgments}
Y. Xu, J. Chang, and T. Shang were financially supported by the Swiss National Science Foundation (SNSF) (Grant No. PP00P2\_179097 and 206021\_139082). L. Das was partially funded through Excellence Scholarship by the Swiss Government. C. J. Yi and Y. G. Shi acknowledged the support from the National Key Research and Development Program of China (No. 2017YFA0302901), and the K. C. Wong Education Foundation (GJTD-2018-01). A. Tiwari received funding from the European Research Council (ERC) under the Marie Sk\l odowska-Curie grant agreement No. 701647. S. S. T. and T. N.  acknowledge support from NCCR MARVEL and from the European Union’s Horizon 2020 research and innovation program (ERC-StG-Neupert-757867-PARATOP). S. S. T. also acknowledges support from the Swiss National Science Foundation (Grant No. PP00P2\_176877). M. S. acknowledges support from NCCR MARVEL. This work was partially supported also by the National Natural Science Foundation of China (Grants No. 11674336 and 11874150).
\end{acknowledgments}

\end{document}